\documentclass[12pt]{arXiv_class}
\usepackage{mathrsfs}

\title{\textbf{A Low-Frequency Tone Sweep Method for in-Service Fault Location in Sub-Carrier Multiplexed Optical Fiber Networks}} 
\author{Gustavo~C.~Amaral,
		Diego~C.~Villafani, 
		Andrea~Baldivieso, 
		Joaquim~Dias~Garcia,\\
		Renata~G.~Leibel,
		Luis E. Y. Herrera,
		Patryk~J.~Urban,
        and~Jean~Pierre~von~der~Weid}

\begin{document}
\maketitle

\begin{abstract}
We demonstrate an optical fiber fault location method based on the frequency response of the modulated fiber optical backscattered signal in a steady state low-frequency step regime. Careful calibration and measurement allows for the reconstruction of the fiber transfer function, which, associated to its mathematical model, is capable of extracting the fiber characteristics. The technique is capable of identifying non-reflective fault events in an optical fiber link and is perfectly compatible with previous methods that focus on the reflective events. The fact that the recuperation of the complex signal is performed in the frequency domain and not via a Fourier Transform enables the measurements to overcome the spatial resolution limitation of Fourier Transform incoherent-OFDR measurements even with frequency sweep ranges down to 100-100000 Hz. This result is backed up by a less than 10 meters difference in fault location when compared to standard OTDR measurements.
\end{abstract}

\section{Introduction}\label{Introduction}

Single-ended optical fiber link characterization techniques have been a point of major focus since the late seventies when Barnosky and others demonstrated that, by measuring the Rayleigh backscattered signal from a pulse sent into the optical fiber, one could extract its physical properties such as attenuation coefficient and eventual power losses along its length \cite{BarnoskiAppOpt1977}. Presently, different techniques for optical fiber characterization are available and standardized such as the Optical Time Domain Reflectometry (OTDR) and Frequency Domain Reflectometry (OFDR) \cite{DericksonBOOK1998}. In any case, embedding any of these fiber-monitoring techniques in Passive Optical Networks (PON) systems remains a challenge in terms of technology and cost-effectiveness not to mention the in-service monitoring paradigm \cite{AmaralPTL2014,ShimOPEX2012}.

OTDR makes use of high-power short optical pulses while measuring the Rayleigh backscattered power as a function of time, its spatial resolution being defined by the duration of the optical pulse. OFDR may present two different flavors: coherent-OFDR (C-OFDR), where the wavelength of a low-power continuous-wave coherent light source is swept and the heterodyne beat between the backscattered light signal and a reference reflection is measured; and incoherent-OFDR (I-OFDR), where a radio-frequency optical sub-carrier frequency is swept and, once again, a heterodyne beat between the detected backscattered signal and the modulating signal is measured. In both C- and I-OFDR, the spatial profile of the fiber is obtained from the Fourier Transform of the detected OFDR signal and the frequency range covered by the optical sweep limits its resolution. A very complete review of frequency domain optical network analysis, where the distinctions between C- and I-OFDR, their pro's, and con's are highlighted, may be found in \cite{jungerman1991frequency}.

In terms of the apparatus necessary to guarantee reasonable high dynamic range and spatial resolution, OTDR and OFDR differ greatly: on one hand, OTDR must employ high bandwidth detection in order to increase spatial resolution since it is directly proportional to the probe pulse width, which, in general, has a trade off associated to a reduction of dynamic range since the energy is reduced along with pulse duration \cite{DericksonBOOK1998, amaral2015automatic}. Frequency domain techniques usually make use of a frequency sweep at the source level and a Fourier transform which provides the conversion from frequency to distance based on the sweeping rate. C-OFDR does not require high frequency response detectors since the detection is performed in the base band, but highly coherent sources are needed in order to achieve long range in optical fibers \cite{vonderWeidJLT1997}. On the other hand, the resolution of the frequency domain techniques, such as in I-OFDR, are given by the frequency sweep range, which imposes high bandwidth detection bringing up again the trade-off between resolution and dynamic range. High resolution is bandwidth demanding whereas high sensitivity imposes low resolution \cite{pierce2000optical, venkatesh1990incoherent}.

Incoherent OFDR measurements have been extensively studied in \cite{pierce2000optical} and Rayleigh scattering has been successfully observed. The mathematical model developed in the time-domain concludes that the amplitude of the Rayleigh scattering signal within an optical sub-carrier is inversely proportional to the sub-carrier's frequency (hereby referred to as the \textit{high-frequency issue}), which complicates signal acquisition when there is a requirement for high spatial resolution. Recently, however, Urban \textit{et al.} \cite{UrbanJLT2016} showed that fiber monitoring can be accomplished in a Sub-Carrier Multiplexed PON (SCM-PON) context by measuring the Rayleigh backscattered signal from a high-frequency (up to 220MHz was reported) inside a 20MHz band, which translates into a 10 meter achievable spatial resolution. The method resembles the step-frequency method presented in the late eighties by Nakayama \textit{et al.} \cite{NakayamaAppOptics1987} however, instead of disregarding the Rayleigh backscattering and focusing on reflective events as in the latter, the effect of the backscattered signal was exploited at its full potential in \cite{UrbanJLT2016}.

An alternative to the \textit{high-frequency issue} of I-OFDR Rayleigh backscattering measurements would be to perform the fiber monitoring in the low-frequency range, up to a few hundred kilohertz, where the $1/f$ hindering effect is not too harsh. The downside of this approach would be the diminished spatial resolution associated to one's measurement when the Fourier Transform of the acquired signal is taken. In order to overcome this preclusion and, at the same time, improve the signal-to-noise ratio at the receiver's end, we tackle the fiber characterization procedure in a different manner: instead of taking the Fourier Transform we attempt to fit the resulting backscattered signal with its associated mathematical model, which, in order to simplify its description, we model as a spatial-dependent phasor and arrive at an equivalent expression to the one found in \cite{pierce2000optical}, but in the frequency domain. By employing a fitting procedure developed in MATLAB$^{\circledR}$, we show that not only a fiber fault's position but also its magnitude can be identified with high precision.

The remainder of the article is laid out as follows. In Section \ref{ExpSet}, we present the experimental apparatus, its characteristics, and limitations. Section \ref{MatModel} attempts to develop an intuitive mathematical model associating the acquired signal to an analytic frequency-dependent function. The relation between a fault's magnitude and the signal's amplitude as well as the limitation associated to eventual reflection events along the fiber are also discussed. The experimental results are presented and discussed in Section \ref{Results}, Section \ref{Impact} presents the technique's limitations and the impact on in-service data transmission, and Section \ref{Conclusion} concludes the article and details our future points of investigation.

\section{Experimental Setup}\label{ExpSet}

\begin{figure}[htbp]
\center
\includegraphics[width=0.7\linewidth]{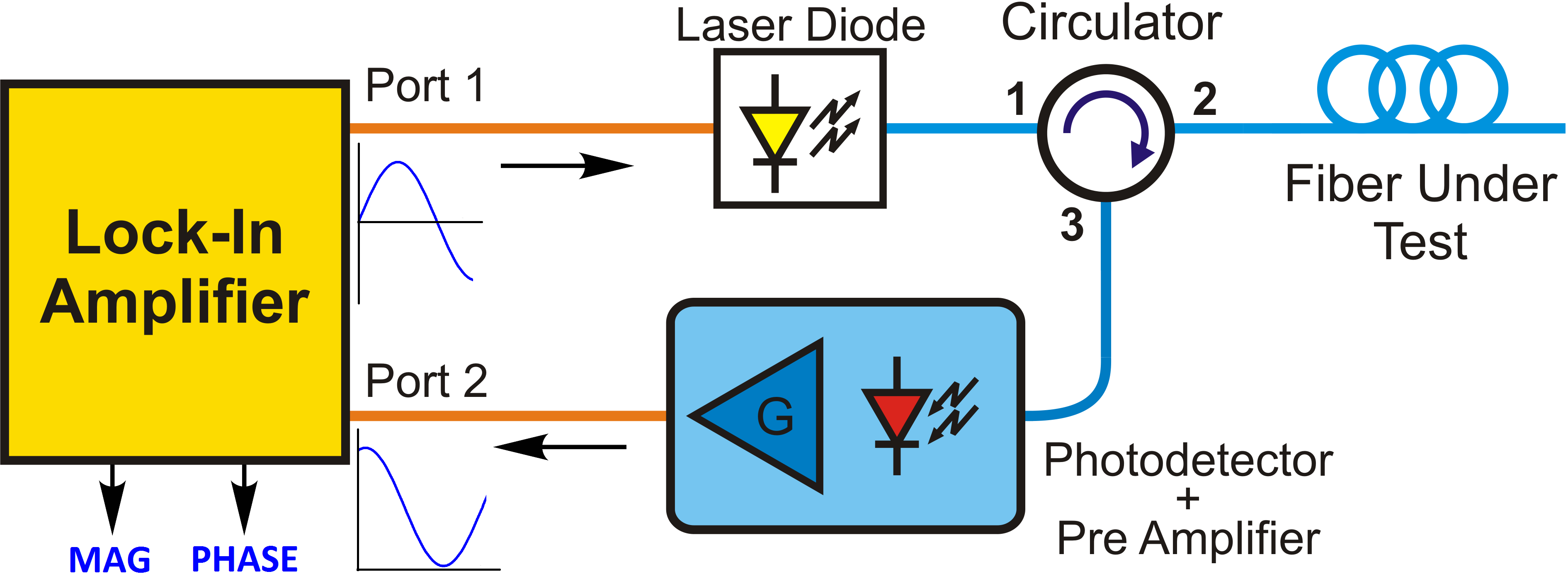}
\caption{Experimental setup of the SCM-PON baseband tone sweep monitoring. The monitoring signal in the base band ($100$ Hz to $100$ kHz) modulates the Laser Diode. The backscattered signal corresponding to the base tone is amplified by the low-frequency photodetector's pre-amplifier and sent to the Lock-In Amplifier.}
\label{fig:Setup}
\end{figure}

The experimental apparatus is rather simple and a block diagram identifying its main structures is presented in Fig. \ref{fig:Setup}. In order to achieve even higher signal-to-noise ratio, we perform the detection in a low-frequency Lock-In Amplifier which is also responsible for generating the modulating optical sub-carrier tone. According to \cite{UrbanJLT2016}, forcing the measuring time to be higher than the round trip time of the sub-carrier signal inside the fiber is equivalent to measuring the steady state response of the system for a given frequency. This way, determining the relative values of magnitude and phase of the backscattered signal for each frequency value in the range is the same as tracing the system's transfer function. By setting the time constant of the Lock-In Amplifier to $300$ milliseconds is sufficient to account for the steady state measurement of any practical optical fiber link available (shorter than $3\cdot10^{4}$ kilometers) and translates into reasonable measuring periods (less than 5 minutes) given that the frequency is swept from 100 Hz to 100 kHz in 100 Hz steps. Furthermore, the sweep time is long enough so that each measurement is acquired after one time constant of the Lock-in Amplifier, which was verified to be short enough to avoid dragging effects on the measured curve. The noise level at each step frequency measurement is $7 \textrm{nVrms}/\sqrt{\textrm{Hz}}$ for the SR530 Analog Lock-Amplifier model employed in the measurements.

The Lock-In Amplifier generates the sweeping tone which falls inside the 100 Hz to 100 kHz band and is responsible for determining the relative amplitude and phase parameters of the backscattered signal which composes the system's transfer function. Note, however, that if we take into account the transfer functions of each element apart from the fiber itself and then proceed to calibrate the resulting signal, we get rid of the contribution from the laser diode, the photodetector, the electronic amplifier, and the Lock In, and are left with the fiber's transfer function. The system's calibration is achieved by inducing a reflective event at port $\#2$ of the optical circulator and measuring the system's response. The laser diode operates linearly within a $30\comm{\mathrm{mA}}$ range centered at $90\comm{\mathrm{mA}}$ bias current. The low-frequency tone with angular frequency $\omega$ ($\omega\!=\!2\pi f$) at full modulation depth has a peak voltage amplitude of $0.75\comm{\mathrm{V}}$ since the laser diode's input impedance is $50\comm{\Omega}$. 

\section{Mathematical Formulation}\label{MatModel}

The optical signal launched into the fiber under test (assumed to have a length $L$) will carry its modulation properties as will the backscattered signal. Upon detection, the modulated backscattered signal will produce an electrical signal with current amplitude proportional to the field amplitude squared and the detector's detectivity \cite{AgrawalBOOK2002}. If we disregard the DC term after the detection, which is a good assumption since we are performing a steady state matched detection at Lock-In Amplifier, the electric signal entering the Lock-In Amplifier as a function of time can be written as $V\pars{t}\!=\!DA\cos\pars{\omega t}$, where $A$ is the detected signal's current amplitude and $D$ is the photodetector's detectivity.

For simplicity purposes, we first neglect possible reflective events throughout our mathematical modeling so the total signal intensity arriving at the Lock-In Amplifier is given solely by the incoherent sum of light scattered from all positions of the fiber. Such reflective events, associated to defective connectors and Fresnel reflections due to an abrupt change in the refractive index, will be input to the model after these first considerations are laid out. If we assume the distribution of scattering centers is uniform along the fiber, we can write the signal entering the Lock-In Amplifier as a spatial-dependent phasor in integral form:
\begin{equation}
S\pars{f} = \intUp{0}{L} C D P_0 e^{j2Kz} dz,
\label{eq:FullFiber_IntForm}
\end{equation}
where $P_0$ is the launched probe signal power at the fiber's input and $C$ is the Rayleigh scattering coefficient, assumed to be constant along the whole fiber length. $K = n\omega / c + j\alpha$ is the complex wave vector of the modulating signal which includes the fiber's attenuation coefficient $\alpha$ in its imaginary part. $n$ is the group index of refraction of the fiber.

We now suppose a faulty \textit{non-reflective} event at position $x$ induces an optical loss $\delta$, where $0\!<\!\delta\!<\!1$. To take into account this fault, we re-write Eq. \ref{eq:FullFiber_IntForm} as:
\begin{equation}
S\pars{f} = \intUp{0}{x} R e^{j2Kz} dz + \delta^2 \intUp{x}{L} R e^{j2Kz} dz,
\label{eq:FiberBreak_IntForm}
\end{equation}
where the first integral corresponds to power being scattered back from the portion of the fiber before the loss event whereas the second integral corresponds to power scattered back from the last portion of the fiber, after the fault event at $x$. The loss factor $\delta$ appears multiplying the second integral twice (hence, squared) due to the fact that light passes two times through the fault in its round trip and we assume that the transmission coefficient is symmetric (it is the same whether light is propagating from the generator to the load or in the opposite direction). To simplify the expression, we condensed the constants into $R = C D P_0$.

Note that the second term in Eq. \ref{eq:FiberBreak_IntForm} can be dismembered into two integrals with different limits
\begin{equation}
\delta^2 \intUp{x}{L} R e^{j2Kz} dz = \delta^2 \intUp{0}{L} R e^{j2Kz} dz - \delta^2 \intUp{0}{x} R e^{j2Kz} dz,
\label{eq:dismember}
\end{equation}
which, when substituted back into Eq. \ref{eq:FiberBreak_IntForm}, yields
\begin{equation}
S\pars{f} = \delta^2 \intUp{0}{L} R e^{j2Kz}dz + \left(1-\delta^2\right)\intUp{0}{x} R e^{j2Kz}dz.
\label{eq:integral_form2}
\end{equation}
The signal arriving at the Lock-In Amplifier can, thus, be interpreted as the combination of two parcels: the modulated backscattered signal from a fiber with length $x$, which contributes with a factor $\delta^2$ to the overall optical power; and the modulated backscattered signal from a fiber with length $L$, which contributes with a factor $\left(1-\delta^2\right)$ to the overall optical power.

Splitting the expression of $S\pars{f}$ into two parcels and dealing with the integrals separately, we find that the results are complex sinusoidal functions with the oscillatory parameter (the product $KL$ which governs the oscillation of the sinusoidal function) being the distance of the fiber:
\begin{align}
\begin{aligned}
S_1\pars{f} &= \delta^2 \intUp{0}{L} R e^{j2Kz}dz = A_1 \frac{\sin\pars{KL}}{K}e^{jKL};\\
S_2\pars{f} &= \pars{1-\delta^2}\intUp{0}{x} R e^{j2Kz}dz = A_2 \frac{\sin\pars{Kx}}{K} e^{jKx},
\end{aligned}
\label{eq:Int2Sinc}
\end{align}
where we have condensed the factors $\delta^2R$ and $\pars{1-\delta^2}R$ into $A_1$ and $A_2$, respectively, and $S\pars{f}\!=\!S_1\pars{f}\!+\!S_2\pars{f}$. We would like to highlight that this modelling approach is rather intuitive as it has a somewhat direct correspondence to OTDR-like traces, i.e., the signal contributions from different portions of the fiber are independently weighted and piled up forming the resulting trace. The difference here is that, instead of step functions as in the standard OTDR, the signal contributions are sinusoidal functions that oscillate more or less within the frequency range of the sweep depending on the fiber length to which they are associated.

In \cite{pierce2000optical}, a similar expression has been derived following a distinct mathematical framework. There, however, the phase factor of the monitoring signal has been disregarded and the authors focused on the magnitude of the phasor in order to obtain an OTDR-like trace by employing a Fourier Transform. As we show in the next section, the phase factor is rather sensitive to changes in the fiber and, if used in conjunction with the magnitude information, may raise the precision of fault location. In order to take both the phase and magnitude informations into account, a fitting procedure in the complex frequency domain is necessary and, although the result is not presented as an OTDR-like trace but, rather, as an event list, it is possible to extend the spatial resolution beyond the limit imposed by the Fourier Transform.

Although the Fourier Transform operator is unambiguous, there are some pitfalls that may arise from computational methods: the non-localizability of the Fourier Transform operator causes it to interpret all frequency components with the same weight; for that reason, there are several methods in the literature that focus on enhancing the localizability of the Fourier Transform operator through signal processing in order to enhance the spatial resolution of frequency measurements (\cite{tropp2010beyond} and references therein). Such methods, however, tend to be very expensive in computational complexity even though the result is equivalent to the fitting procedure demonstrated in this document. The fitting procedure on the frequency domain is much more attractive in our application since the phase information is highly dependent on the position of the break which translates into a computationally cheap and accurate method for determining the position of a fault thus enhancing the achievable spatial resolution with respect to a straightforward Fourier Transform.

The result conveyed by Eq. \ref{eq:Int2Sinc} is that $S\pars{f}$ is the sum of two sinusoidal terms which take different arguments corresponding to the length of fiber traversed by the backscattered optical signal. From the expressions of Eq. \ref{eq:Int2Sinc}, we may attempt to write a generic term $S_g$ for the backscattered signal from a fiber of arbitrary length $X$. This generic term is a function of both the frequency and the fiber length and has the form presented below, where we took the liberty of expanding and simplifying all terms:
\begin{equation}
S_g\pars{f,X} = \frac{e^{j\frac{4\pi f n}{c}X} e^{-2\alpha X}-1}{j\frac{4\pi f n}{c}-2\alpha}.
\label{eq:genericTerm}
\end{equation}
Note that, for an arbitrary number of non-reflective faults along the fiber link, the procedure presented in Eq. \ref{eq:dismember} can be used indefinitely. Moreover, the backscattered signal from an optical fiber link which contains a number $N$ of fault events along its length will be represented as the sum of $N$ generic terms presented in Eq. \ref{eq:genericTerm}, each dependent on the fault position $X_i$ and with different amplitude factor $A_i$, so we may write:
\begin{equation}
S\pars{f} = \sum_{i=1}^{N}A_i S_g\pars{f,X_i}.
\label{eq:S_final}
\end{equation}

The expression of $A_i$ gets progressively intricate as the number of fault event rises as can be perceived from the right-hand side of Eq. \ref{eq:integral_form2}: with $N$ terms, for instance, the value of $A_N$ depends on the values of all $A_{i}$ from $N-1$ downto $1$ which also holds true for $A_{N-1}$ and so forth. Extending the expression of $A_i$ for any number of fault events yields the following:
\begin{equation}
A_i = \pars{\prod_{n=0}^{i-1}\delta_n^2}\pars{1-\delta_i^2}.
\label{eq:A_i}
\end{equation}

Once the non-reflective events have been considered and modeled, we turn our attention to the parcel associated to reflective events. Observing reflections along an optical fiber link has been a matter of extensive research and, as previously commented, Nakayama \textit{et al.} were able to detect such reflections with a similar experimental apparatus \cite{NakayamaAppOptics1987}. In a way, therefore, reflection events are manageable and the method hereby proposed offers a means of not only identifying these specific events, but also the non-reflective ones. As we know, reflective events are always accompanied by a loss of intensity due to energy conservation conditions (namely, the $r^2+t^2=1$ condition at any partially reflective interface) but the converse does not hold, i.e., an intensity loss may or may not be accompanied by a reflection. Generally, loss events that are not associated to a reflection peak in standard OTDR measurements are due to absorption or irradiation effects, such as macro- and micro-bend losses \cite{herrera2016investigation}.

Reflection events are more easily observed since the \textit{high-frequency issue} does not arise in modeling its amplitude and a sinusoidal function which reflected back will arrive at the Lock-In Amplifier as a sinusoidal function with lesser magnitude and constant phase. Hence, taking reflective events into account is as straightforward as including a constant magnitude and constant phase factor to the model. We, then, rewrite Eq. \ref{eq:S_final} as
\begin{equation}
S\pars{f} = \sum_{i \in I}A_i S_g\pars{f,X_i} + \sum_{\ell \in I_R}B_{\ell}e^{jKX_{\ell}}
\label{eq:S_final_reflec}
\end{equation}
with $I_R \subseteq I$, meaning that the set of reflective events is a subset of all the events but not necessarily equal to the latter.

In the same fashion as Nakayama \textit{et al.} considered the Rayleigh backscattering some sort of noisy contribution that could spoil the reflection measurements performed \cite{NakayamaAppOptics1987}, we consider the reflection events as noisy contributions that can spoil our fiber characterization method. Suppose, for instance, that a optical fiber link presents a non-reflective event at position $x$ and a reflective fiber-end at position $L$. If the reflective fiber-end contribution rises above the backscattering power level, the non-reflective event won't be observed even though it is there.

\section{Experimental Results}\label{Results}

\begin{figure}[htbp]
\center
\includegraphics[width=0.7\linewidth]{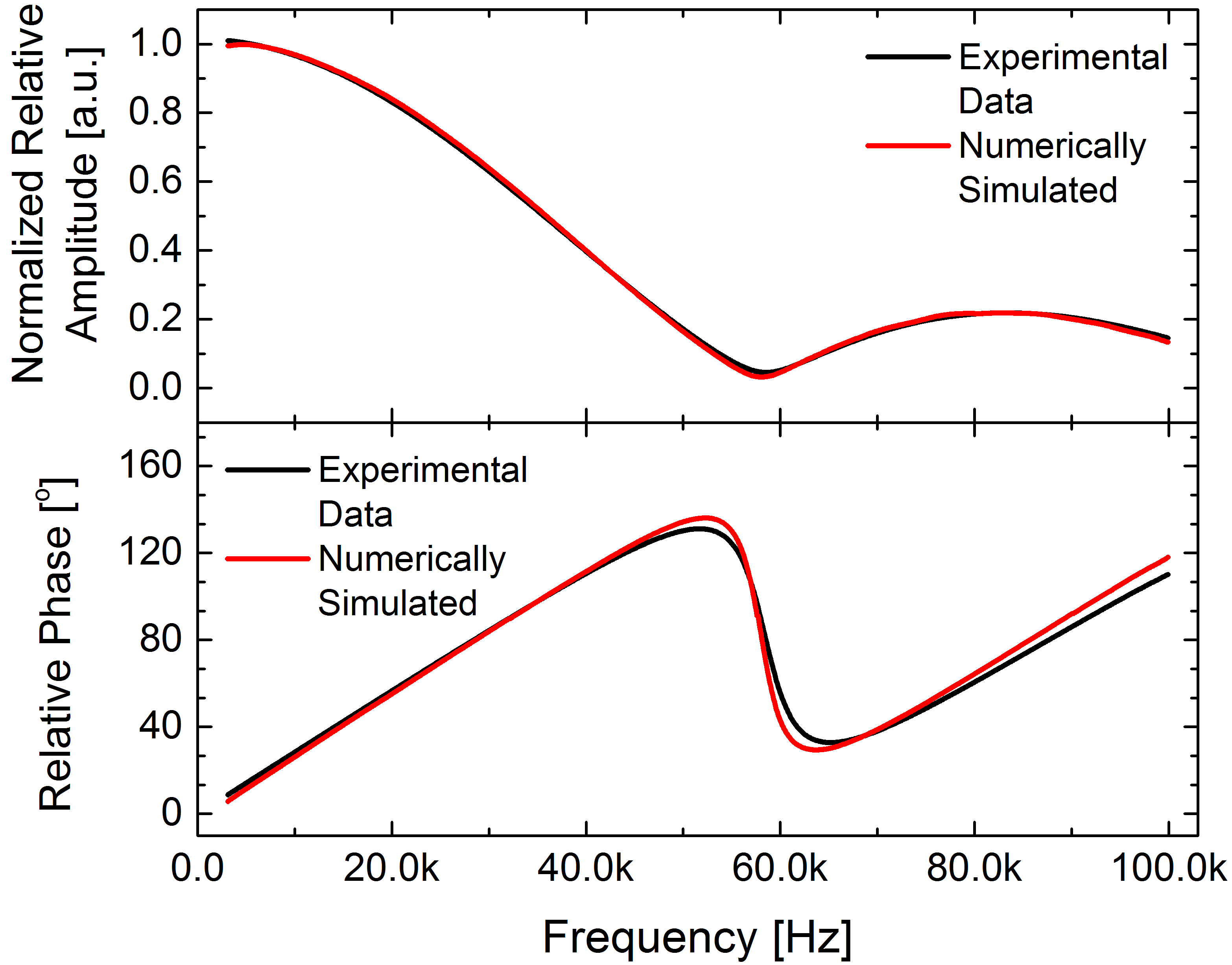}
\caption{Simulation versus experimental results of a $1.55$-km standard optical fiber. a) Amplitude data series; b) Phase data series.}
\label{fig:AmpPhaseSimulationComp}
\end{figure}

We first wish to validate the mathematical modelling of the signal arriving at the Lock-In Amplifier. For that, we simulated the results from Eq. \ref{eq:S_final} using $N\!=\!1$, $X\!=\!1.55$km, and a frequency range of $100$Hz to $100$kHz. We then compared the amplitude and phase data with the results acquired from measuring a $1.55$-km standard optical fiber within the same frequency range using our proposed method. The amplitude of the modulating signal throughout all experiments were kept at one eighth of the full modulation depth. The result is depicted in Fig. \ref{fig:AmpPhaseSimulationComp}.

Even though the results show good agreement between simulated and experimentally acquired data, there is still a slight discrepancy, specially in the phase data series. Note, however, that the generic term presented in Eq. \ref{eq:genericTerm} depends on $f$ and $X$ but also on $\alpha$ and $n$, the fiber's attenuation coefficient and the group index of refraction of the fiber, respectively. Since, from an optical link monitoring perspective, there is usually no \textit{a priori} access to such information, we utilized standard values within the International Telecommunication Union (ITU-T) specification range, i.e., $\alpha\!=\!0.21$dB/km and $n\!=\!1.4682$ \cite{ITUT_G652}. Nevertheless, these values may suffer variations from one fiber to the other and the discrepancy perceived between simulated and experimental data may be attributed to that.

Observe that, from a systematic point of view, if one was capable of identifying which frequency-dependent sinusoidal functions are present in a data series originated by the monitoring procedure described -- which is equivalent to finding all the generic terms of Eq. \ref{eq:genericTerm} that compose the signal --, the eventual positions of faults could be determined. Before that, and in order to further investigate the correspondence between the mathematical model and the experimentally acquired data, as well as determining the availability of this fiber characterization method, we conducted laboratory tests referenced by a standard OTDR device with previously unknown fibers (unknown length and attenuation coefficient). First, the amplitude and phase profiles were extracted from the fiber in its original configuration using the Lock-In Amplifier. Afterwards, a fiber bend loss was induced in an arbitrary position also with unknown magnitude. The goal is to identify how a fiber bend alters the transfer function's profile, so the amplitude and phase profiles of the modified fiber were extracted using the Lock-In Amplifier.

\begin{figure}[htbp]
\center
\includegraphics[width=0.7\linewidth]{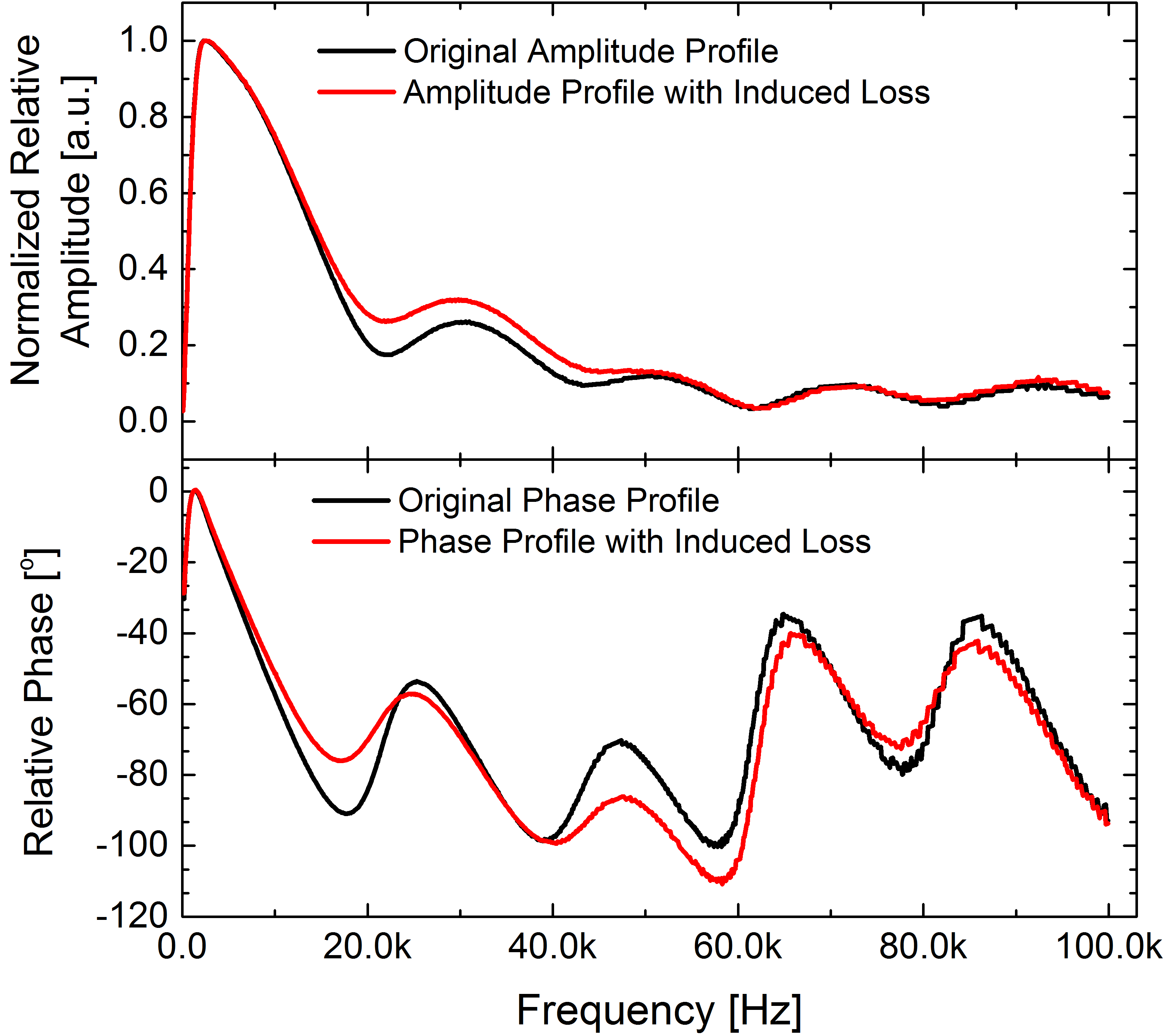}
\caption{Amplitude profile (a) and  phase profile (b) of the low-frequency sweep technique as measured in the output of the Lock-In Amplifier. Tempering with the fiber original structure by inducing an extra loss has a direct effect on the measured signal.}
\label{fig:AmpPhaseComp}
\end{figure}

From the results of Fig. \ref{fig:AmpPhaseComp}-a and Fig. \ref{fig:AmpPhaseComp}-b, it is clear that tempering with the fiber has a direct effect on the signal measured with the Lock-In Amplifier, so the possibility of quantifying the effect by employing our knowledge of the mathematical model is a straightforward assumption.

To assess the capability of identifying a fault event along an optical fiber link, the OTDR trace of the unknown fiber was acquired in its original and fault-induced conditions. At the same time, a model fit of the amplitude and phase profiles was ran based on an \textit{extensive search} heuristic developed in MATLAB$^\circledR$: the algorithm sorts an arbitrary fiber length and calculates the generic component (Eq. \ref{eq:genericTerm}) of our mathematical model; it then calculates the minimum square error between the original profile (the fiber without the induced fault) summed with this generic component and the profile after the induced fault; the position ($X_i$), amplitude factor ($A_i$), and attenuation coefficient parameters are adjusted to find the candidate that best fits the new result.

Note that the nature of the mathematical model enables one to test an arbitrary number of generic components in order to best estimate the fiber fault position. In a sense, the spatial resolution is limited by the amount of generic components one is inclined to test since the higher this number, the longer the algorithm will take to come up with a result. In the examples that follow, the generic components were created each separated by 1 meter, i.e., for a 4-km fiber, four thousand components will be tested and the one that best fits the acquired signal will be selected. The non-linear search algorithm employed in MATLAB$^{\circledR}$, however, must deal with four parameters at a time (position, magnitude, $\alpha$, and $n$) so the average processing time revolves around $100$ milliseconds. Increasing this separation over 1 meter will be computationally costly and ever slower, for which reasons we set 1 meter as our default separation between generic components achieving a $\sim\!5$ minutes computing time. The developed algorithm along with a testbench data series is available at \cite{GitAlgorithm}.

We tested two different optical links: Link $\#1$ has its amplitude versus phase profile presented in Fig. \ref{fig:AmpPhaseComp} and corresponds to a $4.90$-km long composition of two fibers with the induced fault at $\sim1.58$ kilometers. Link $\#2$ corresponds to a $6.67$-km composition of two fibers with the induced fault at $\sim3.36$ kilometers. The results are shown in Table \ref{table:Comp1}, for which we found a less than $10$ meters difference between the standard OTDR and the low-frequency sweep measurements in both cases. This results are a clear indication that the Fourier Transform limitation of I-OFDR has indeed been overcome since a 100kHz frequency sweep would allow for a maximum spatial resolution of 2 kilometers.

\begin{table}[htbp]
\centering
\caption{Standard OTDR vs Low-Frequency Sweep}
\begin{tabular}{c c c c c}
 & \multicolumn{2}{c}{Standard OTDR} & \multicolumn{2}{c}{Low-Frequency Sweep} \\
\cmidrule(lr){2-3} \cmidrule(lr){4-5}
           & Pos [km]  & Mag [dB] & Pos [km]  & Mag [dB]\\
Link $\#1$ & 1.587     & 1        & 1.582     & 1.05    \\
Link $\#2$ & 3.363     & 0.5      & 3.370     & 0.6     \\
\end{tabular}
\label{table:Comp1}
\end{table}

Since the links were composed by different fibers spliced together, the mean attenuation coefficient $\alpha$ for both links was also determined neglecting its variation from one fiber to another. This assumption was necessary in order to avoid too many free parameters in the fit, improving the convergence and robustness of the method. In other words, the value of $\alpha$ found by the software corresponds to a mean of the attenuation coefficient along the whole optical link. Link $\#1$ presented an estimate of $\alpha = 0.21$ [dB/km] whereas Link $\#2$ presented an estimate of $\alpha = 0.20$ [dB/km], values within the expected range for single mode standard optical fibers. Calculation based on the standard OTDR trace for both optical links indicate a good correspondence of these values, with $\alpha = 0.209$ and $\alpha = 0.203$ for both Link $\#1$ and $\#2$ respectively.

\begin{figure}[htbp]
\center
\includegraphics[width=0.7\linewidth]{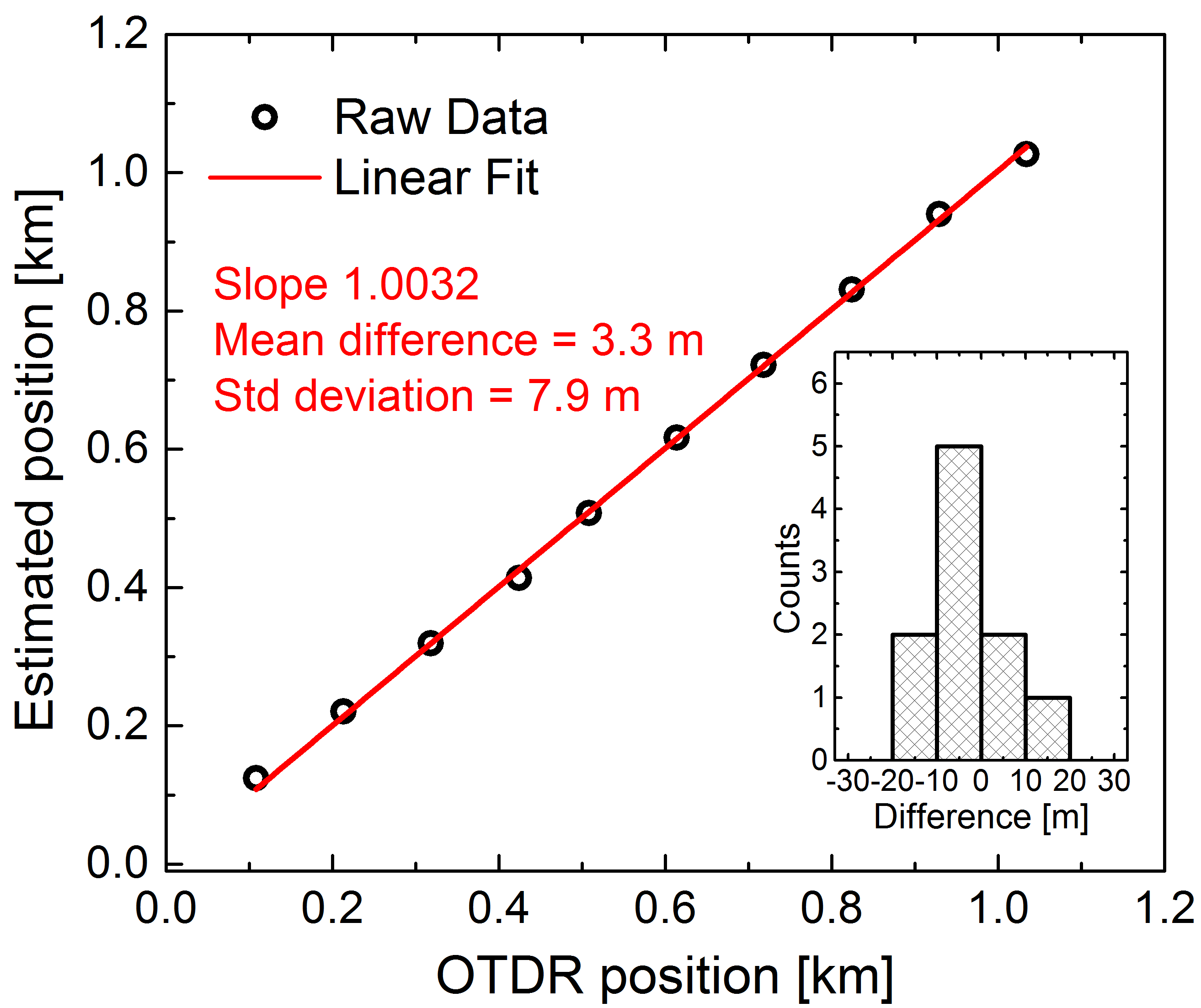}
\caption{Estimated position of the non-reflective induced fault in the $\sim1.10$-km using the low-frequency sweep and standard OTDR methods. The inset details the histogram of the estimated position error with no more than $20$ meters errors. The results show that the low-frequency method can accurately determine the position of a new fault on an optical fiber link.}
\label{fig:EstimateVSReal}
\end{figure}

Another striking result of the low-frequency fiber characterization is presented in Fig. \ref{fig:EstimateVSReal}. A $\sim\!1.10$-km fiber link was previously characterized and, subsequently, faults were induced -- one at a time --  at different positions. Each fault was measured by the low-frequency technique and also characterized by a standard OTDR. The results are plotted against each other in Fig. \ref{fig:EstimateVSReal} and not only attest that the model describing the Rayleigh backscatter signal is robust, but also indicates that fiber monitoring could be achieved using the low-frequency step method.

The results of Fig. \ref{fig:EstimateVSReal} are acquired by creating a 1.1-km optical fiber link with 0.1-km standard single-mode fiber stretches spliced together. The frequency response of the fiber without faults is determined via the experimental procedure previously described. Then, a single fiber bend is induced every 100 meter, and the fault fiber frequency response is acquired for each different fault position. The position of the fault is determined by the extensive search algorithm. The reference measurements for each faulty fiber is acquired via a standard OTDR measurement. Small refractive index differences between the spliced fibers composing the link and the algorithm's lack of precision are the responsible for the discrepancy between reference and measured fault positions. This discrepancy is translated on the deviation of the linear fit's slope.

\section{Limitations and Impact on Data Transmission}\label{Impact}

To evaluate the limitations of the monitoring method, the achievable dynamic range was investigated. Since the output signal is not a standard OTDR trace, a different methodology had to be developed in order to determine the dynamic range: a loss with increasing magnitude was induced in a 1.55-km optical fiber and the resulting backscatter profile was processed by the developed algorithm. The induced loss was also characterized by a standard OTDR, which acted as a reference value. Table \ref{table:Comp2}, shows the intensity of the induced real loss (OTDR measurement) which varies from a clean link to a link with a loss of $6.9$ dB in steps of $\sim\!2$ dB. Note that the contribution of the break component varies from $0\%$ (clean link) to $100\%$ (maximum achievable dynamic range).

\begin{table}[htbp]
\centering
\caption{Low-Frequency Tone Sweep Monitoring Dynamic Range}
\begin{tabular}{c c c c}
Real Loss      & Loss Component      & Fiber End           \\
$\bract{dB}$   & Contribution [$\%$] & Contribution [$\%$] \\
0              & 0.00                & 100.00              \\
1.50           & 43.15               & 56.85               \\
2.60           & 67.64               & 31.35               \\
4.70           & 77.48               & 22.51               \\
6.20           & 97.52               & 2.48                \\
6.90           & 100.00              & 0.00                \\
\end{tabular}
\label{table:Comp2}
\end{table}

The fiber end contribution varies inversely, as expected. According to these results, the dynamic range is limited to a maximum of $\sim\!6.2$ dB, i.e., the maximum attenuation throughout the link which still permits one to identify the fiber end. In the access systems of RoF, the optical link distribution does not, usually, exceed 15 km. Therefore, even though the technique achieves low dynamic range, it is an interesting candidate for short range mobile fronthaul applications with low optical splitting ratios \cite{UrbanJLT2016}.

The possibility of in-service network monitoring has been assessed by measuring the resulting Error Vector Magnitude (EVM$_{rms}$) for different data signal amplitude under two conditions: with; and without the monitoring tone. Once again, the monitoring tone amplitude was kept at one eighth of the laser's full modulation depth in order to simulate an eight-channel SCM network. The result indicates that the monitoring signal deteriorates the quality of the communication, even if by a very small factor. Under these conditions, however, the required EVM$_{rms}$ for LTE communication is respected with a large margin \cite{EVM}.

\begin{figure}[htbp]
\center
\includegraphics[width=0.7\linewidth]{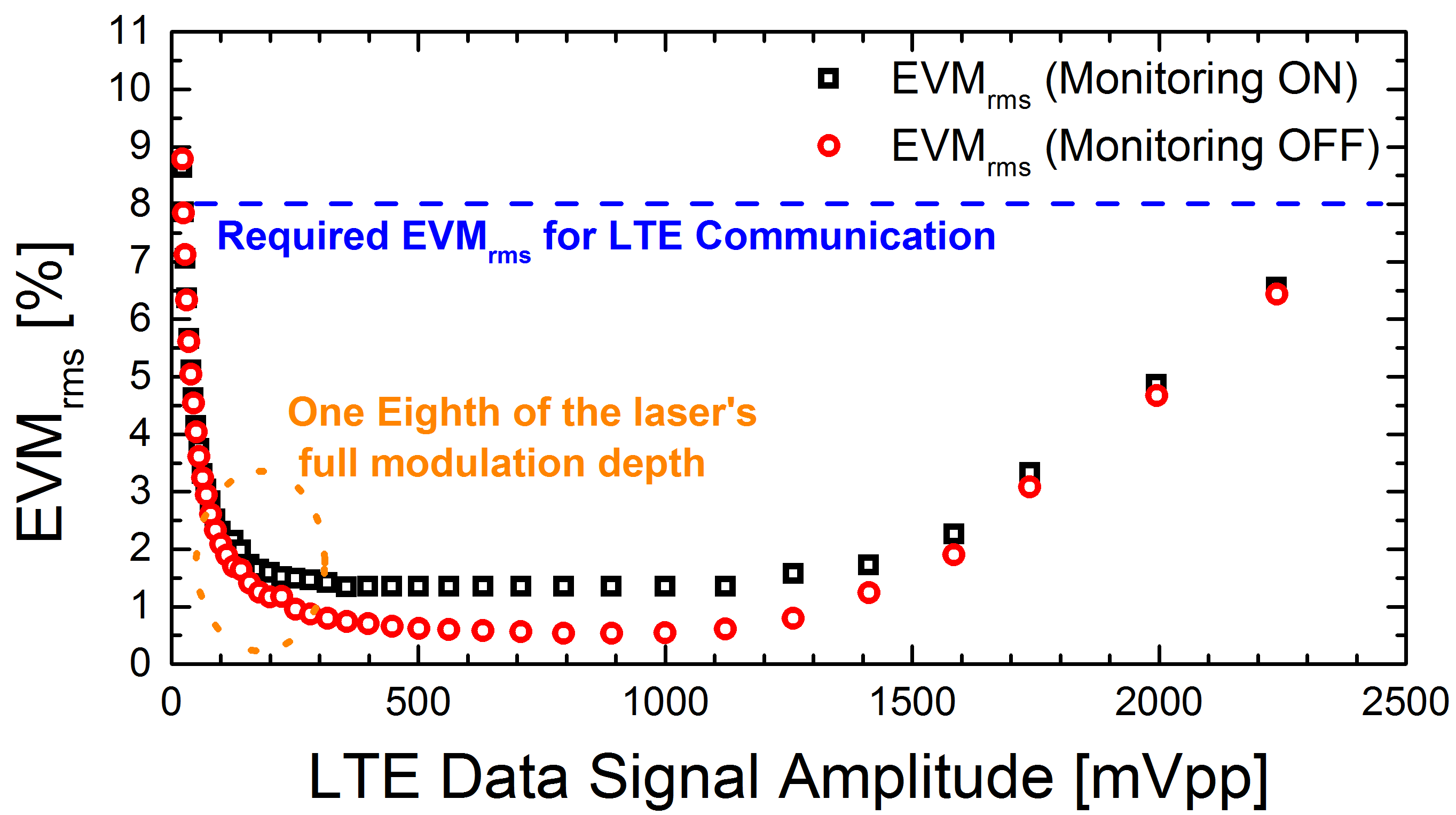}
\caption{Error Vector Magnitude of LTE 64-QAM communication when the monitoring tone is turned on and off.}
\label{fig:InServiceMonitoring}
\end{figure}

It can be observed that, above a certain value of the data signal's amplitude, the quality of communication is deteriorated. This behavior is associated to the laser's non-linear response region, achieved when the modulation signal violates the laser's full modulation depth at around 1.5 [Vpp], or 30 [mA].

\section{Conclusions}\label{Conclusion}

The recent interest in the deployment of SCM-PON and cost restrictions imposed by modern commercial needs bring up low cost embedded monitoring techniques as an essential tool for operational success of access networks. Although a method capable of replacing one optical sub-carrier channel by a monitoring probe has been presented in \cite{UrbanJLT2016}, dynamic range is limited by the 1/f factor impairing backscattering measurements inside an optical sub-carrier. The results presented here shows that an embedded in-service short-reach link monitoring can be performed and the impact over data transmission is expected to be not higher than the cross contribution from any two sub-carrier channels due to the probe signal's very nature. Using the baseband for monitoring purposes, the method benefits of the high-sensitivity low-frequency detection and further overcomes the \textit{high-frequency issue}, fully preserving the RF sub-carrier band for data transmission. Fulfilling the steady state condition over the round trip path along the fiber during the measurement is imperative for the performance of the technique and a feature that distinguishes it from the technique presented in \cite{NakayamaAppOptics1987}. Nevertheless, minute-long measurement times are achieved in fibers a few kilometers long.

The mathematical model presented here is similar to the one previously developed in \cite{pierce2000optical} but with a different approach, namely the development in the frequency domain when one represents the incoming signal as an integral of phasors originated from the modulated light coming from distinct portions of the fiber. The intuitiveness of the model is useful since the signal is not transformed back to the spatial domain as in standard C- or I-OFDR and the results are given in the form of an event table as opposed to an OTDR-like trace. The algorithm developed to treat the signal, even though simple and still limited enables highly reliable fault location, as backed by our experimental results contrasted to standard OTDR measurements. This is possible due to the fact that several signal component candidates may be tested and the one that best approximates the acquired complex signal in the frequency domain is selected. Using the phase combined with the magnitude information of the backscattered signal, we could overcome the Fourier Transform spatial resolution limitation of low-frequency measurements.

We believe that, as the method developed by \cite{UrbanJLT2016}, the method presented here may also figure as a good candidate for embedded in-service short-reach SCM-PON monitoring, not demanding a sub-carrier channel for that purpose. Assuming that the data channels are delegated to highest frequencies (in the order of hundreds of megahertz), the base band would be unoccupied and could be reserved for the low-frequency probing tone herewith described. We also believe that the standard deviation reported in Figure \ref{fig:EstimateVSReal} is the limitation of localizability of the phase measurement even when the fitting procedure in the frequency domain is employed. Even though we surpass the standard Fourier Transform limitation, there is still a Fourier Transform limitation to the technique, i.e., we cannot retrieve perfect information of the fiber unless an infinite frequency span is performed. Unfortunately, the Lock-In Amplifier employed in the measurements is incapable of reaching higher than $100$ kHz frequencies and, therefore, we are limited to this spatial resolution of $\sim\!10$ meters. Higher frequency spans should improve the resolution, so conducting the experiments with a device specified for higher frequencies is a goal for future developments.

Location of multiple faults is, unfortunately, still impracticable with the present signal processing technique. We believe, however, that it could be achieved and leave it as our major future point of investigation.

\section*{Acknowledgment}
The Innovation Centre, Ericsson Telecomunica\c{c}\~{o}es S.A., Brazil and brazilian agencies CNPq and FAPERJ supported this work.

\bibliographystyle{IEEEtran}

\end{document}